# Dynamics of Neutrino Wave Packet in the Tachyon-like Dirac Equation

Luca Nanni

Faculty of Natural Science, University of Ferrara, 44122 Ferrara, Italy

luca.nanni@student.unife.it

**Abstract** In this study the tachyon-like Dirac equation, formulated by Chodos to describe superluminal neutrino, is solved. The analytical solutions are Gaussian wave packets obtained using the envelope method. It is shown that the superluminal neutrino behaves like a pseudo-tachyon, namely a particle with subluminal velocity and pure imaginary mass that fulfils the energy-momentum relation typical of classical tachyons. The obtained results are used to prove that the trembling motion of the particle position around the median, known as Zitterbewegung, also takes place for the superluminal neutrino, even if the oscillation velocity is always lower than the speed of light. Finally, the pseudo-tachyon wave packet is used to calculate the probability of oscillation between mass states, obtaining a formula analogous to the one obtained for the ordinary neutrino. This suggest that in the experiments concerning neutrino oscillation is not possible to distinguish tachyonic neutrinos from ordinary ones.

## 1. Introduction

Recently, the results obtained from the experiments dedicated to the measurement of neutrino masses and the relative oscillation phenomena [1-9], often interpretable assuming a superluminal behaviour of the particle, have led many physicists to enhance their efforts towards the formulation of new theories beyond the Standard Model [10-13]. In this framework, the pioneering works by Surdashan, Feinberg and Recami [14-17] on the physics of tachyons were taken into consideration with the aim of formulating a field theory consistent with the theory of relativity extended to superluminal motions. One of these theories is that of Chodos [18], whose governing equation is obtained from the Tanaka Lagrangian [19]. This Lagrangian reads:

$$\mathcal{L}_t = i\bar{\psi}_t \gamma^5 \gamma^\mu \partial_\mu \psi_t - \mu_t \bar{\psi}_t \psi_t \qquad (1)$$

and holds for half-integer spin tachyons. In Eq. (1) $\bar{\psi}_t = \psi_t^\dagger \gamma^0$, $\gamma^5 = i\gamma^0\gamma^1\gamma^2\gamma^3$, $(\gamma^5)^2 = \mathbb{1}$, $\mu$ is the tachyonic mass and $\hbar = c = 1$. The subscript *t* refers to tachyon and will be used in all quantities introduced below. It remains clear that $\psi_t$ and $\bar{\psi}_t$ are treated as independent variables. Moreover, the Dirac matrix $\gamma^5$ is related to the fifth current and its presence in the Lagrangian (1) proves the chiral nature of the particles it describes. The Lagrangian (1) is Hermitian and fulfils the classical tachyonic energy-momentum relation $E^2 = p^2c^2 - \mu^2c^4$. By inserting this Lagrangian in the Euler-Lagrange equation [20], the Chodos equation is recovered:

$$[i\gamma^5\gamma^\mu\partial_\mu - \mu_t]\psi_t = 0 \qquad (2)$$

Eq. (2) is a tachyon-like Dirac equation, one of the most used to study half-integer spin superluminal particle. Jentschura proved that this equation is *CP* and *T* invariant, but the associated Hamiltonian operator is not Hermitian and loses parity symmetry [21]. However, the Hamiltonian fulfils the symmetry properties of a pseudo-Hermitian operator.

In this study, the Chodos equation is solved, in terms of Gaussian wave packets with positive and negative frequencies, for a tachyon propagating along the *z* direction. In this way we obtain the equations of the envelope functions which, once solved, show that the group velocity is always subluminal. Therefore, the tachyonic neutrino described by relativistic quantum mechanics is a pseudo-tachyon, namely a particle that fulfils the energy-momentum relation typical of an imaginary mass, but which propagates at subluminal velocity, being the group velocity equal to that of the quantum particle [22]. This result was obtained also by Salesi using a different tachyon-like Dirac- equation and following a different approach [23]. It is therefore necessary to distinguish the meaning between the group velocity, which finds its natural place in the quantum study of the tachyonic neutrino, and the classical velocity of the particle which, in principle, is not upper bound. The theory developed in this work, however, shows that the two velocities are related to each other and that the tachyon velocity is upper bound. This is an indirect proof that tachyonic neutrinos, in the picture of relativistic quantum mechanics, are unstable particles that decay following mechanisms, already investigated in other works, in order to return in the subluminal realm.

As occurs for a relativistic particle with half-integer spin in the Dirac equation, the Zitterbewegung phenomenon [24], represented by the rapid oscillation of the position of the particle with respect to the median of the Gaussian packet, also takes place for the tachyonic neutrino. This effect is due to the fact that, although the wave packets with positive and negative frequencies are orthogonal, once inserted in the integral $\langle\psi^-|z|\psi^+\rangle$, which represents the average position of the particle along *z* direction, show a non-vanishing overlap which leads to the typical interference of the Zitterbewegung. It is also proved that, unlike what happens for a relativistic Dirac particle, the oscillation velocity of Zitterbewegung is always lower than the speed of light.

The wave packet approach is used to calculate the probability of neutrino oscillation [25]. In this study we apply the obtained Gaussian packet to calculate the oscillation probability of tachyonic neutrino. This need the assumption that even in the tachyonic regime the neutrino may oscillate between possible mass states. We will show that the formula of probability oscillation is analogous to that expected for the ordinary neutrino. This is a confirmation that the (pseudo)- tachyonic neutrino cannot be distinguished from the ordinary one in the experiments concerning oscillation.

## 2. Tachyonic Wave Packets

Let us consider a superluminal neutrino propagating along $z$ direction. For classical physics, the particle velocity $u_t$ can take any value higher than the speed of light. Before proceeding, we clarify that the mass $\mu_t$ and the tachyonic Lorentz factor $\gamma_t = [(1 - u_t^2/c^2)]^{-1/2}$ are pure imaginary, being $\gamma_t = -i|\gamma_t|$. Therefore, the product $\mu_t \gamma_t$ is always real, as well as the momentum $p$ and the energy $E$ [26]. The mass energy $\mu_t c^2$ is instead pure imaginary. The Chodos equation for this model reads:

$$[i\hbar\gamma^5\gamma^0\partial_t - i\hbar c\gamma^5\gamma^3\partial_z - \mu_t c^2]\psi_t = 0 \tag{3}$$

Using the gamma Dirac matrices, the operators $\gamma^5\gamma^0$ and $\gamma^5\gamma^3$ are:

$$\gamma^5\gamma^0 = \begin{pmatrix} 0 & 0 & \bar{1} & 0 \\ 0 & 0 & 0 & \bar{1} \\ 1 & 0 & 0 & 0 \\ 0 & 1 & 0 & 0 \end{pmatrix} \quad ; \quad \gamma^5\gamma^3 = \begin{pmatrix} \bar{1} & 0 & 0 & 0 \\ 0 & 1 & 0 & 0 \\ 1 & 0 & 1 & 0 \\ 0 & 1 & 0 & \bar{1} \end{pmatrix} \tag{4}$$

where $\bar{1}$ means $-1$. The bispinor $\psi_t$ has two components, $\psi^+$ and $\psi^-$, each of which is associated with the positive and negative frequencies (energies):

$$\begin{cases} \psi^+ = \begin{pmatrix} u_1^+ \\ u_2^+ \end{pmatrix} \exp\{i(kz - \omega^+ t)\} \\ \psi^- = \begin{pmatrix} u_1^- \\ u_2^- \end{pmatrix} \exp\{-i(kz - \omega^+ t)\} \end{cases} \tag{5}$$

where $k = p/\hbar$ and $\omega^+ = E^+/\hbar$, while $p$ is the $z$-components of four-momentum. Introducing the bispinor $\psi_t$ in the Eq. (3) we get a system of four linear differential equations:

$$\begin{pmatrix} i\hbar c\partial_z - \mu_t c^2 & 0 & -i\hbar\partial_t & 0 \\ 0 & -i\hbar c\partial_z - \mu_t c^2 & 0 & -i\hbar\partial_t \\ i\hbar\partial_t & 0 & -i\hbar c\partial_z - \mu_t c^2 & 0 \\ 0 & i\hbar\partial_t & 0 & i\hbar c\partial_z - \mu_t c^2 \end{pmatrix} \begin{pmatrix} u_1^+ \\ u_2^+ \\ u_1^- \\ u_2^- \end{pmatrix} e^{\pm i(kz - \omega^+ t)} = 0$$

We note that the matrix in the left-side of the equation is anti-Hermitian. As it is known, anti-Hermitian operators are the infinitesimal generators of unitary transformations and in quantum mechanics are associated to imaginary eigenvalues [27]. The system can be easily solved giving us all the spinor components:

$$\begin{cases} u_1^+ = -E/(pc + \mu_t c^2) \quad ; \quad u_2^+ = E/(pc - \mu_t c^2) \\ u_1^- = -E/(pc - \mu_t c^2) \quad ; \quad u_2^- = -E/(pc + \mu_t c^2) \end{cases} \tag{6}$$

Considering that $E = \gamma_t \mu_t c^2$ and $p = \gamma_t \mu_t c$, Eqs. (6) can be written as functions of the dimensionless factor $\gamma_t$:

$$\begin{cases} u_1^+ = |\gamma_t|/(|\gamma_t|+1) \; ; & u_2^+ = -|\gamma_t|/(|\gamma_t|-1) \\ u_1^- = |\gamma_t|/(|\gamma_t|-1) \; ; & u_2^- = |\gamma_t|/(|\gamma_t|+1) \end{cases} \quad (7)$$

We see that the two-component spinors are real and orthogonal. The obtained solutions must be normalized by applying the usual normalization procedure:

$$\int \psi^\dagger \psi = 1 \quad \Rightarrow \quad \mathcal{N} = \sqrt{2(|\gamma_t|^2+1)}/(|\gamma_t|^2-1)$$

where $\mathcal{N}$ is the normalization factor. Therefore, the plane waves solutions of the Chodos equation can be summarized as:

$$\psi^\pm = \frac{\sqrt{2(|\gamma_t|^2+1)}}{(|\gamma_t|^2-1)} \begin{pmatrix} \frac{|\gamma_t|}{(|\gamma_t|\pm 1)} \\ \mp \frac{\gamma_t}{(|\gamma_t|\mp 1)} \end{pmatrix} exp\{\pm i(kz - \omega^+ t)\} \quad (8)$$

The trend of the real spinor components $u_1^+$ and $u_2^+$ is shown in Fig. 1:

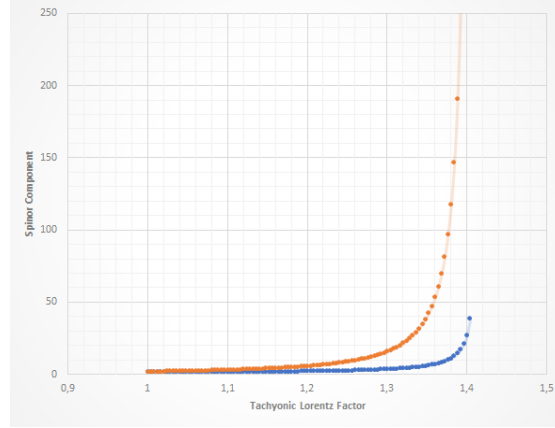

**Figure 1**: spinor components $u_1^+$ (blue line) and $u_2^+$ (orange line) vs tachyonic Lorentz factor.

As can be seen, the two components converge to zero as $u_t \to c$, while as $u_t \to \sqrt{2}c$ thay tend to separate e diverge towards infinity with different slopes.

For the models we are developing, we want the solutions to be Gaussian wave packets. Therefore, we need to find an envelope function that multiplied by the tachyonic plane wave provides the expected wave packet. This function is a smooth curve outlining the extremes in amplitude of the rapidly varying single wavefunction that spreads in space and time. Its profile must be that typical of a Gaussian function. To do this we have to set a given value of the classical velocity of the tachyonic neutrino, denoted by $u_0$, to which correspond the wave vector $k_0$ and the angular frequency $\omega_0$. These values represent the centre of the Gaussian packet. The Gaussian spinor can be written as:

$$\psi_G^\pm = \frac{\sqrt{2(|\gamma_t|^2+1)}}{(|\gamma_t|^2-1)} f^\pm(t,z) \begin{pmatrix} \frac{|\gamma_t|}{(|\gamma_t|\pm 1)} \\ \mp \frac{\gamma_t}{(|\gamma_t|\mp 1)} \end{pmatrix} exp\{\pm i(k_0 z - \omega_0^+ t)\} \quad (9)$$

where $f^\pm(t,z)$ are the envelop functions for positive and negative frequencies. Introducing function (9) in the Eq. (2) we get the two differential equations that, once solved, provide the explicit form of the envelope functions:

$$\begin{cases} \left(\dfrac{\partial}{\partial t} - c\dfrac{u_1^+}{u_1^-}\bigg|_{\gamma_0}\dfrac{\partial}{\partial z} + \dfrac{\Lambda_0^+}{i\hbar u_1^-}\bigg|_{\gamma_0}\right)f^+(t,z) = 0 \\ \left(\dfrac{\partial}{\partial t} + c\dfrac{u_2^-}{u_2^+}\bigg|_{\gamma_0}\dfrac{\partial}{\partial z} + \dfrac{\Lambda_0^-}{i\hbar u_2^+}\bigg|_{\gamma_0}\right)f^-(t,z) = 0 \end{cases} \quad (10)$$

where $\gamma_0$ is the module of the tachyonic Lorentz factor corresponding to the velocity $u_0$ and $\Lambda_0^\pm = 2\mu_t c^2 u_1^\pm\big|_{\gamma_0}$. The numerical coefficient of the second term in Eqs. (10) is the propagation velocity, which coincides with the group velocity of the wave packet. Using Eq. (7) we obtain the explicit form of these velocities:

$$c\dfrac{u_1^+}{u_1^-}\bigg|_{\gamma_0} = -c\dfrac{u_2^-}{u_2^+}\bigg|_{\gamma_0} = \left(\dfrac{\gamma_0 - 1}{\gamma_0 + 1}\right)c \leq c \quad \forall u_o > c \quad (11)$$

Eq. (11) proves that the neutrino described by Chodos equation behaves like a pseudo-tachyon, namely a particle propagating with subluminal velocity $v = c(\gamma_0 - 1)/(\gamma_0 + 1)$ but fulfilling the energy-momentum relation of a tachyon. This result, which may seem surprising and unexpected, was also obtained by Salesi following a different approach [23]. It must be clear that the velocity $v$ in Eq. (11) is of quantum mechanics nature and is obtained trough the action of the operator $i\hbar c\partial_z$, that is conserved under the action of Lorentz transformations. Eq. (11) gives the relation between the pseudo-tachyon velocity and the classical velocity $u_0$.

To solve Eqs. (10) it is sufficient to impose that the envelope functions $f^\pm$ are Gaussian. As a basic function we can take the following Gaussian envelope, widely used in quantum optics [28-29]:

$$f^\pm(t,z) = \left[\dfrac{1}{\sqrt{2\pi(\sigma + iAt)}}\right]^{1/2} exp\left\{-\dfrac{(z \mp v_0 t)^2}{2\sigma(\sigma + iAt)}\right\} \quad (12)$$

Where $v_0$ is the pseudo-tachyon velocity given by Eq. (11), $\sigma$ is the wave packet dispersion coefficient along $z$ direction and $A$ is a numerical constant that must be found. Introducing function (12) in the first equation of (10) and calculating its value in the point $(z,t) = (0,0)$, we get an algebraical equation from which constant $A$ is easily obtained:

$$A^+ = -\omega_0\theta_0 \quad where \; \theta_0 = \Lambda_0^+/(\hbar\omega_0 u_1^-|_{\gamma_0}) \quad (13)$$

Eq. (13) holds for positive frequencies. Repeating the same procedure for the second equation of (10) we obtain the value of the constant $A$ for the negative frequencies:

$$A^- = \omega_0\theta_0 \quad where \; \theta_0 = \Lambda_0^+/(\hbar\omega_0 u_1^-|_{\gamma_0}) \quad (14)$$

Therefore, the Gaussian envelope function for positive and negative frequencies is:

$$f^{\pm}(t,z) = \left[\frac{1}{\sqrt{2\pi(\sigma \mp i\omega_0\theta_0 t)}}\right]^{1/2} exp\left\{-\frac{(z \mp v_0 t)^2}{2\sigma(\sigma \mp i\omega_0\theta_0 t)}\right\} \qquad (15)$$

The term $\omega_0\theta_0$ is the tachyonic correction to dispersion of the wave packet. In Fig. 2 are shown the real and imaginary components of the wave packet:

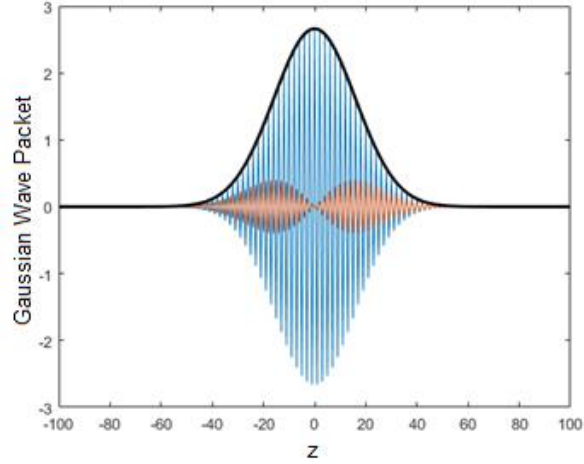

**Figure 2**: real (blue area) and imaginary (red area) components
Of pseudo-tachyon wave packet (in arbitrary units).

Let us analyse in detail this term replacing to $\omega_0$ and $\theta_0$ their explicit forms:

$$\omega_0\theta_0 = \frac{E_0}{\hbar}\frac{2|\mu_t|c^2}{\hbar\omega_0}\frac{u_1^+}{u_1^-}\bigg|_{\gamma_0} = 2\omega_{Plank}\left(\frac{\gamma_0-1}{\gamma_0+1}\right) \qquad (16)$$

where $\omega_{Plank}$ is the angular frequency given by $|\mu_t|c^2/\hbar$. For the factor $(\gamma_0 - 1)/(\gamma_0 + 1)$, which is simply the relativistic factor $\beta = v/c$ of the pseudo-tachyon, the following limits hold:

$$\lim_{u_0 \to c}\left(\frac{\gamma_0-1}{\gamma_0+1}\right) = 1 \ ; \ \lim_{u_0 \to \sqrt{2}c}\left(\frac{\gamma_0-1}{\gamma_0+1}\right) = 0 \ \vdots \ \lim_{u_0 \to \infty}\left(\frac{\gamma_0-1}{\gamma_0+1}\right) = -1 \qquad (17)$$

Since the pseudo-tachyon has a subluminal velocity, the factor $(\gamma_0 - 1)/(\gamma_0 + 1)$ must range between [0,1]. This means that the classical velocity of neutrino is upper bound to $u_0 = \sqrt{2}c$. This is a further confirmation of how quantum physics can lead to completely different results from those obtained applying classical physics, even in the tachyon field. We conclude this section noting that as $u_0$ increases the tachyonic dispersion of the wave packet progressively decreases up to a minimum value of zero, corresponding to the classical velocity $\sqrt{2}c$. This behaviour is similar to that of a wave packet associated with an ordinary relativistic particle, where the dispersion correction factor is proportional to $1/\gamma^3$ [29].

### 3. Zitterbewegung of Pseudo-tachyon Neutrino

A fermion that obeys the Dirac equation presents a rapid oscillation of the position along the direction of propagation, known as Zitterbewegung [24]. This oscillation is due to the interference between states with positive and negative energy and occurs with a frequency of $2\omega_{Plank}$. Assuming that the motion takes place in the $z$ direction, the equation of the position of the ordinary particle around the median is:

$$z(t) = \frac{pc^2}{E}t + \frac{1}{2}i\frac{\hbar c}{E}\left[v(0) - \frac{pc}{E}\right][exp\{-2i\omega_0 t\} - 1] \qquad (18)$$

The first term of Eq. (18) is the particle position along the direction of propagation, while the second one is the oscillation due to the partial overlap of the positive and negative frequency wave packets. We want to investigate whether this behaviour also takes place for the pseudo-tachyon wave packet. To do this we follow the same approach used by Park to study the Zitterbewegung of a relativistic electron wave packet [29]. In this regard, we consider the wave function $\psi(t,z)$ given by the linear combination of the two wave packets given by Eq. (9):

$$\psi(t,z) = c_1\psi_G^+ + c_2\psi_G^- \quad c_1, c_2 \epsilon \Re \qquad (19)$$

The mean value of particle position is given by $\langle\psi|z|\psi\rangle$. The real coefficients $c_1$ and $c_2$ are such as to ensure that the function $\psi(t,z)$ is normalized:

$$\{c_1, c_2 \epsilon \Re \ : \ \langle\psi|\psi\rangle = 1\}$$

Although the wave packets $\psi_G^+$ and $\psi_G^-$ are orthogonal, the integral $\langle\psi|z|\psi\rangle$ is non vanishing. The solution of this integral, that for the relativistic electron has been already obtained by Park [29], is:

$$\langle z \rangle = v_o t(c_1^2 - c_2^2) +$$
$$+ \left[1 + \frac{1}{4\sigma^2}\left(\frac{\hbar}{2\gamma_0^2|\mu_t|c}\right)^2\right]^{-1}\left(\frac{\hbar}{2\gamma_0^2|\mu_t|c}\right)\left(\frac{\sigma^2}{\sigma^2+(\omega_0\theta_0 t)^2}\right)^{1/4} exp\left\{-\frac{(v_0 t)^2}{2[\sigma^2+(\omega_0\theta_0 t)^2]}\right\}\cdot$$
$$\cdot \sin\left[2\omega_0 t - \frac{(\omega_0\theta_0 t)^2}{\sigma^2+(\omega_0\theta_0 t)^2}(\gamma_0^2 - 1)\omega_0 t + \varphi/2\right]\cdot 2c_1 c_2$$

where $\varphi = arctg(\omega_0\theta_0 t)$ and $v_0 = c(\gamma_0 - 1)/(\gamma_0 + 1)$. The first term in the right-side is the median position of the particle modulated by the coefficient $(c_1^2 - c_2^2)$, which can be positive or negative. Therefore, the median position does not coincide with the centre of the wave packet. The product between the first and the second factor in the right-size represent the maximum oscillation amplitude, while the product between the third factor and the exponential represent the oscillation damping term. Since it has been shown that the term $\omega_0\theta_0$ ranges within $[0, 2\omega_{Plank}]$, the damping $\Sigma$ is maximum when $u_0 = \sqrt{2}c$:

$$\Sigma|_{u_0=\sqrt{2}c} = \left(\frac{\sigma^2}{\sigma^2+(2\omega_{Plank}t)^2}\right)^{1/4} exp\left\{-\frac{(v_0 t)^2}{2[\sigma^2+(2\omega_{Plank}t)^2]}\right\} \qquad (20)$$

We note also that when $u_t \to c$, i.e. $\gamma \to \infty$, the maximum oscillation amplitude $I$ goes quickly to zero, while it increases progressively as $u_t \to \sqrt{2}c$ up to the upper limit given by:

$$I|_{u_0=\sqrt{2}c} = \frac{\lambda_{Plank}}{2}\left(1 + \frac{\lambda_{Plank}^2}{16\sigma^2}\right)^{-1} \qquad (21)$$

where $\lambda_{Plank}$ is the reduced Plank wave length of pseudo-tachyon neutrino. Finally, we observe that when $u_t \to \sqrt{2}c$ the coefficient $\omega_0 \theta_0$ vanish and $\langle z \rangle$ becomes:

$$\langle z \rangle|_{u_0=\sqrt{2}c} = v_o t(c_1^2 - c_2^2) + 2\left[1 + \frac{1}{4\sigma^2}\left(\frac{\hbar}{2|\mu_t|c}\right)^2\right]^{-1} \left(\frac{\hbar}{2|\mu_t|c}\right) exp\left\{-\frac{(v_0 t)^2}{2\sigma^2}\right\} sin[2\omega_0 t] \cdot 2c_1 c_2$$

The oscillation frequency around the median position is the argument of *sin* function; Taylor expanding this function around $t \to 0$ and truncating the sum at the first term we get:

$$\omega_{ZB} \cong \omega_0 \left(2 - \frac{(\omega_0 \theta_0 t)^2}{\sigma^2}(\gamma_0^2 - 1)\right) \tag{22}$$

Using eqs. (21) and (22) we obtain the Zitterbewegung velocity:

$$v_{ZB} = I|_{u_0=\sqrt{2}c} \omega_{ZB} = c\left[1 - \frac{(2\omega_{Plank} t)^2}{\sigma^2}\left(\frac{\gamma_0 - 1}{\gamma_0 + 1}\right)(\gamma_0^2 - 1)\right] \tag{23}$$

From Eq. (23) we see that, in the range $[c, \sqrt{2}c]$, the velocity $v_{ZB}$ is always lower than the speed of light.

## 4. Oscillation of Pseudo-tachyon Neutrino

Recently, Caban et al. have shown that the hypothesis of tachyonic neutrino leads to the same oscillation phenomenon of ordinary neutrino [30]. This result can be used to validate the theory presented in section 2. To do this we use the wave packet approach [25] considering that we have to write down the function $\psi_G^{\pm}$ for each mass eigenstate. By limiting the attention to the only the positive frequency and assuming that there are only three mass eigenstates, the evolved state of the pseudo-tachyon neutrino produced in the initial state $v_\alpha$ is:

$$|v(t,z)\rangle = \sum_{i=1}^{3} U_{\alpha i}^* \psi_{Gi}^+ |v_i(t,z)\rangle \tag{24}$$

where $U_{\alpha i}$ is the leptonic mixing matrix that we suppose holds also for tachyonic particles. The oscillation probability from a state of imaginary mass $\mu_i$ to a state with mass $\mu_\beta$ is:

$$P(v_\alpha \to v_\beta) = \left|\langle v_\beta | v(t,z)\rangle\right|^2 = \left|\langle v_\beta | \sum_{i=1}^{3} U_{\alpha i}^* \psi_{Gi}^+ |v_i(t,z)\rangle\right|^2 \tag{25}$$

To solve this integral one must know the phase difference between the IN and OUT states:

$$\Delta \Phi = \Delta E \cdot t - \Delta p \cdot z \tag{26}$$

Considering that we are in an ultra-relativistic regime (where $u_t \to c$ and $(\gamma_0 - 1)/(\gamma_0 + 1) \cong 1$) and that we are dealing with a wave packet, the following approximations hold:

$$\Delta E \ll E \quad and \quad \Delta p \cong \sigma_p = \hbar/\sigma$$

Therefore, $\Delta E$ can be Taylor expanded obtaining:

$$\Delta E \cong \frac{\partial E}{\partial p}\sigma_p + \frac{\partial E}{\partial \mu^2}\Delta\mu^2 = v\sigma_p - \frac{c^4}{2E}\Delta\mu^2 \qquad (27)$$

where $v$ is the pseudo-tachyon velocity obtained in section 2 and $E$ is the tachyonic energy-momentum relation. Supposing that the error $\sigma_p$ affecting the momentum is of the order of $p_0$, then the wave packet dispersion $\sigma$ can be reworked as follows:

$$\Delta p \cong \sigma_p = \frac{\hbar}{\sigma} \Rightarrow \sigma = \frac{\hbar}{p_0} = \frac{\hbar}{|\mu_t|u_0} = \frac{\hbar}{|\mu_t|c}\frac{\gamma_0-1}{\gamma_0+1} = \lambda_{Plank}\frac{\gamma_0-1}{\gamma_0+1}$$

Substituting this result in Eq. (27) we get:

$$\Delta E \cong \frac{\hbar c}{\lambda_{Plank}}\left(\frac{\gamma_0-1}{\gamma_0+1}\right)^2 - \frac{\Delta\mu^2 c^4}{2E} \qquad (28)$$

and substituting Eq. (28) in Eq. (26) we obtain the explicit form of $\Delta\Phi$:

$$\Delta\Phi = -\left[L - \left(\frac{\gamma_0-1}{\gamma_0+1}\right)ct\right]\left(\frac{\gamma_0-1}{\gamma_0+1}\right)\frac{\hbar}{\lambda_{Plank}} - \frac{\Delta\mu^2 c^4}{2E}t \qquad (29)$$

For the oscillation to take place there must be interference between the mass states and this is possible only if the term $L - (\gamma_0 - 1)/(\gamma_0 - 1)$ is of the order of dispersion $\sigma$. But this means that the first term of Eq. (29) is $\ll 1$ and can be neglected. Therefore:

$$\Delta\Phi \cong -\frac{\Delta\mu^2 c^4}{2E}t \qquad (30)$$

Considering that $t \cong L/c$ (since we are in ultra-relativistic limit) we arrive at the final result:

$$\Delta\Phi \cong -\frac{\Delta\mu^2 c^4}{2E}\frac{L}{c} = -\frac{\Delta\mu^2 2}{2p}L \qquad (31)$$

Therefore, the oscillation probability for a tachyonic neutrino in the Chodos equation is:

$$P(\nu_\alpha \to \nu_\beta) = \left|\sum_{i=1}^{3} U_{\alpha i} \exp\left\{i\frac{\Delta\mu^2 2}{2p}L\right\} U_{\alpha i}^*\right|^2 \qquad (32)$$

Eq. (32) is analogous, except for the sign of the square mass, to the oscillation probability expected for ordinary neutrino [31]. Therefore, in the state of the art of current experiments concerning the phenomenon of oscillation, is not possible to distinguish the bradyonic or tachyon nature of the neutrino. This confirm the result obtained by Caban et al. [30] and prove the correctness of the theory developed on the Chodos equation.

## 6. Discussion

In this study, the Chodos equation for the tachyonic neutrino has been solved, obtaining Gaussian wave packets, with positive and negative frequencies, analogous to those obtained by solving the ordinary Dirac equation [32]. The equations obtained for the envelope functions, which guarantee the Gaussian shape of the wave packet, show that the group velocity is always subluminal. Since the group velocity coincide with that of neutrino propagation, one concludes that a particle with a half-integer spin with a classical velocity greater than the speed of light, in the framework of quantum mechanics behaves like a subluminal fermion with imaginary mass. Furthermore, the theory shows that the results retain their physical meaning if the classical analogue of the tachyonic velocity is upper bound by $\sqrt{2}c$. This suggests that such particles are theoretically possible but are unstable and do not decrease their energy by increasing their velocity, as the classical tachyon theory would predict [26]. This hypothesis has been deeply investigated by Jentschura [33], who proposed possible mechanisms of decay.

The first validation test of the obtained solutions is represented by the study of the Zitterbewegung effect. The theory shows that this effect also occurs for the Chodos equation, highlighting the typical oscillation of the position of the particle around the median. However, the oscillation velocity always remains lower than the speed of light, unlike what was predicted for the electron by the Dirac equation, where this velocity results equal to the speed of light [34].

The second validation test of the theory is represented by the calculation of the oscillation probability of tachyonic neutrino, assuming that the modules of the imaginary masses are identical to those of the three ordinary neutrinos and that the leptonic mixing matrix is the same as the current model. Following the wave packet approach, it is obtained a probability formula having the same form of that used for ordinary neutrino, confirming the same result achieved by other authors [30] using a different approach, which provides for the impossibility of distinguishing tachyonic and ordinary neutrinos in the oscillation phenomena.

This work proves that the equation proposed by Chodos for the description of superluminal neutrino is consistent with what is expected from a theory that has its foundation in the Dirac equation. Many physical-mathematical aspects of Dirac equation also recur for that of Chodos [21] and its application to real problems reproduce results obtained by following other approaches [30]. The most evident result, however, is that which proves that in the Chodos equation the neutrino behaves like a subluminal particle that obey the energy-momentum relationship typical of classical tachyons. This could be one of the reasons why, to date, there is no experimental evidence that proves with certainty the possible tachyonic nature of neutrinos and that the efforts to detect them must be oriented towards the precision measurement of the value of their square masses. Only negative values of this quantity can confirm whether or not the neutrinos can have imaginary mass states.